# Hybrid Cavity from Tunable Coupling between Anapole and Fabry-Pérot Resonance or Anti-resonance

Aoning Luo[1], Haitao Li[1], Ken Qin[1], Jingwen Ma[2], Shijie Kang[1], Jiayu Fan[1], Yiyi Yao[1], Xiexuan Zhang[1], Jiusi Yu[1], Boyang Qu[1], Xiaoxiao Wu[1*]

[1]*Modern Matter Laboratory and Advanced Materials Thrust, The Hong Kong University of Science and Technology (Guangzhou), Nansha, Guangzhou 511400, China*

[2]*Department of Electrical and Electronic Engineering, The University of Hong Kong, Hong Kong, China*

*xiaoxiaowu@hkust-gz.edu.cn

## Abstract

Enhancing light-matter interactions depends critically on the ability to tailor photonic modes at subwavelength scales, and combining distinct resonant modes has shown remarkable potential unattainable by individual resonances alone. Despite recent advances in anapole metasurfaces for energy confinement and Fabry-Pérot (FP) cavities for spectral control, their synergistic coupling and resulting opportunities remain largely unexplored due to challenges such as precise nanoscale assembly in experiments. Here, we demonstrate that embedding a terahertz (THz) anapole metasurface within a tunable FP cavity results in a hybrid cavity that demonstrates exotic properties as the anapole transitions between coupling to FP resonances and anti-resonances via cavity-length tuning. At room temperature, we observe ultrastrong coupling (> 30% of the anapole frequency) between anapoles and FP resonances, generating tunable-dispersion polaritons that blend favorable properties of both modes. Meanwhile, anapole spectrally aligns with FP anti-resonances, leading to weak coupling that narrows anapole's transmission peak linewidth by one order of magnitude and enhances its local density of states (LDOS) near the metasurface correspondingly. With exceptional capabilities including formation of polaritons and significant enhancement of LDOS, the hybrid cavity enables strong interaction with functional materials, paving the way for exploration of quantum optics, molecular sensing, and ultrafast nonlinear photonics.

# 1 Introduction

Over the past decade, metasurfaces have emerged as a powerful platform for investigating photonic coupling due to their exceptional versatility[1–9]. A key advance is the anapole metasurface, where destructive interference between the far-field radiation of electric and toroidal dipoles enables subwavelength electromagnetic energy confinement[10–12], advancing light-matter interaction studies[13–16]. Meanwhile, FP cavities with broad spectral tunability provide a new degree of freedom to manipulate the surrounding photonic environment[17], facilitating selective excitation of optical modes that align with collective excitations (e.g., phonons[18,19] or excitons[20–22]). These platforms have been successfully coupled with quantum dots[23,8], transition metal dichalcogenides[24,25], and organic molecules[26] across a wide frequency range, achieving applications like Bose-Einstein condensation[27–29], laser technology[30,31], quantum information processing[32], nonlinear optics[33], and polaritonic chemistry[34]. However, the coupling between anapole metasurfaces and FP cavities remains experimentally unexplored[35–37], despite growing interest in hybrid metasurface-FP cavity systems[38–42]. This lack is primarily due to challenges in achieving precise alignment at micrometer and shorter wavelengths, as well as generating collimated, directional beams at centimeter and longer wavelengths. The absence of such a comprehensive understanding of their synergistic potential, especially in experiments, significantly impedes the research and development of advanced photonic devices leveraging their combined strengths.

Here, we propose, theoretically analyze, and experimentally demonstrate the coupling between an anapole metasurface and an FP cavity, which leads to a hybrid cavity with exotic properties. The metallic metasurface transforms FP modes of the full cavity into resonances or anti-resonances of separated cavities, depending on its placement at the standing-wave nodes or anti-nodes of the full cavity, respectively. This functionality establishes position control as an additional degree of freedom for reconfigurable photonic coupling, enabling on-demand switching between their strong and weak coupling regimes. Taking the typical case of metasurface positioning at the center of the full cavity, metasurface-mediated coupling between two separated cavity modes that coalesce into the full cavity’s near-degenerate modes with distinctive parity. With the match of energy levels between cavity and anapole resonances, we construct a three-order coupled oscillator model (COM) and transfer matrices to analyze mode coupling. Due to the spatial parity, we can control over anapole-symmetric cavity resonance interaction, allowing us to decouple the anti-symmetric FP mode from mode hybridization. Moreover, strongly coherent energy exchange induces a considerable Rabi splitting, which causes the hybrid cavity to enter the ultrastrong coupling regime. The generated polaritonic branches inherit subwavelength mode confinement from anapole and dispersion tunability from the cavity mode, achieving concurrent control of resonant frequency and mode volume. On the other hand, in the weak coupling regime, we report the first observation of single-band linewidth narrowing induced by the spectral alignment of the anapole with FP anti-resonances (in contrast to previous studies focusing on resonances[22,43–45]). This phenomenon arises from anomalous Purcell enhancement and cavity transparency, leading to strongly localized anapole and directly enhancing LDOS. Our work also establishes a direct correlation between the linewidth compression ratio and LDOS spectra. These findings demonstrate that the hybrid cavity offers an original and practical paradigm for spectrally and spatially tailoring LDOS enhancement, serving as a promising platform for exploring light-matter interactions, especially at long wavelengths, where the coupling strength is inherently weak.

## 2 Results and Discussion

### 2.1 Structure and Design Principles

We consider a THz metasurface integrated within an FP cavity, as shown in Fig. 1(a). The metallic metasurface comprises a periodic array of split-ring slots (lattice constant $a$ = 0.5 mm) fabricated on a 0.13-mm-thick substrate (relative permittivity $\varepsilon_r$ = 2.2). The FP cavity is formed by two dielectric slabs ($\varepsilon_r$ = 6.5), and each slab has a deep-subwavelength thickness $t_{slab}$ = 0.13 mm (< $\lambda_{slab}/10$, $\lambda_{slab}$ being wavelength in the slab), effectively suppressing interference effects within the slab[46] (see Note 1 for sample characterization). To elucidate the mode evolution and coupling mechanism, we decompose the hybrid cavity and analyze each component first. In the analysis, the metal is approximated as a perfect electric conductor (PEC). For the bare metasurface, the normalized electric (|**E**|) and magnetic (|**H**|) field distributions of its lowest-order eigenmode reveal an out-of-phase relationship between electric dipole (**P**) and toroidal dipole moments ($ik$**T**) [Fig. 1(b)], both oriented along the $y$-direction. These distributions, consistent with multipole decomposition and volume-splitting secondary multipole analysis[47] (see Note S2 in Supporting Information for details), confirm the presence of a non-radiative anapole. Meanwhile, the FP cavity supports leaky modes due to standing waves formed between the two slabs [see Fig. S4(b) in Note S3], with photon energy levels denoted as $E_i = \hbar\omega_i$. If separated by a continuous metallic layer in the middle, it will suppress odd-order FP modes of the original full cavity into anti-resonance of the half cavities, while sustaining resonance for the even-order FP modes (see Note S3 in Supporting Information). In this configuration, two near-degenerate eigenmodes $FP_A$ and $FP_B$ will emerge at energy levels corresponding to even-order FP modes [Fig. 1(c)]. In fact, their linear combination $(|FP_A\rangle \pm |FP_B\rangle)/\sqrt{2}$ produces two cavity modes with distinct symmetries that interact with the anapole. Despite a substrate-induced perturbation breaking their perfect degeneracy, the mirror-symmetry approximation remains effective for our key analysis (see Note S4 in Supporting Information for verification). For clarity, we first consider the lowest-order energy levels. Within the hybridization framework [Fig. 1(d)], anapole strongly couples to FP resonances at zero energy detuning ($\delta = \hbar\omega_{anapole} - \hbar\omega_2$) as the coupling strength $g$ exceeds overall dissipation $\gamma = (\gamma_{anapole} + \gamma_{FP})/2$. In this scenario, strong coupling induces Rabi splitting ($\Omega_{Rabi} = 2g = \omega_+ - \omega_-$), reorganizing energy levels into three distinct eigenfrequencies: the polaritonic lower branch (LB) and upper branch (UB), along with the uncoupled middle branch (MB)[17]. When the normalized coupling ratio $g/\omega_{\mathrm{anapole}}$ approaches and exceeds 0.1, the hybrid cavity enters the ultrastrong coupling regime[48]. On the other hand, weak coupling stems from the anapole mode overlapping with the FP anti-resonance ($\hbar\omega_{\mathrm{anapole}} = \hbar\omega_1$), yielding a narrowed linewidth without an obvious frequency shift [Fig. 1(e)]. Therefore, precise adjustment of the FP cavity length tailors the photonic environment around the metasurface correspondingly through selective excitation and coupling with FP resonances or anti-resonances at targeted frequencies.

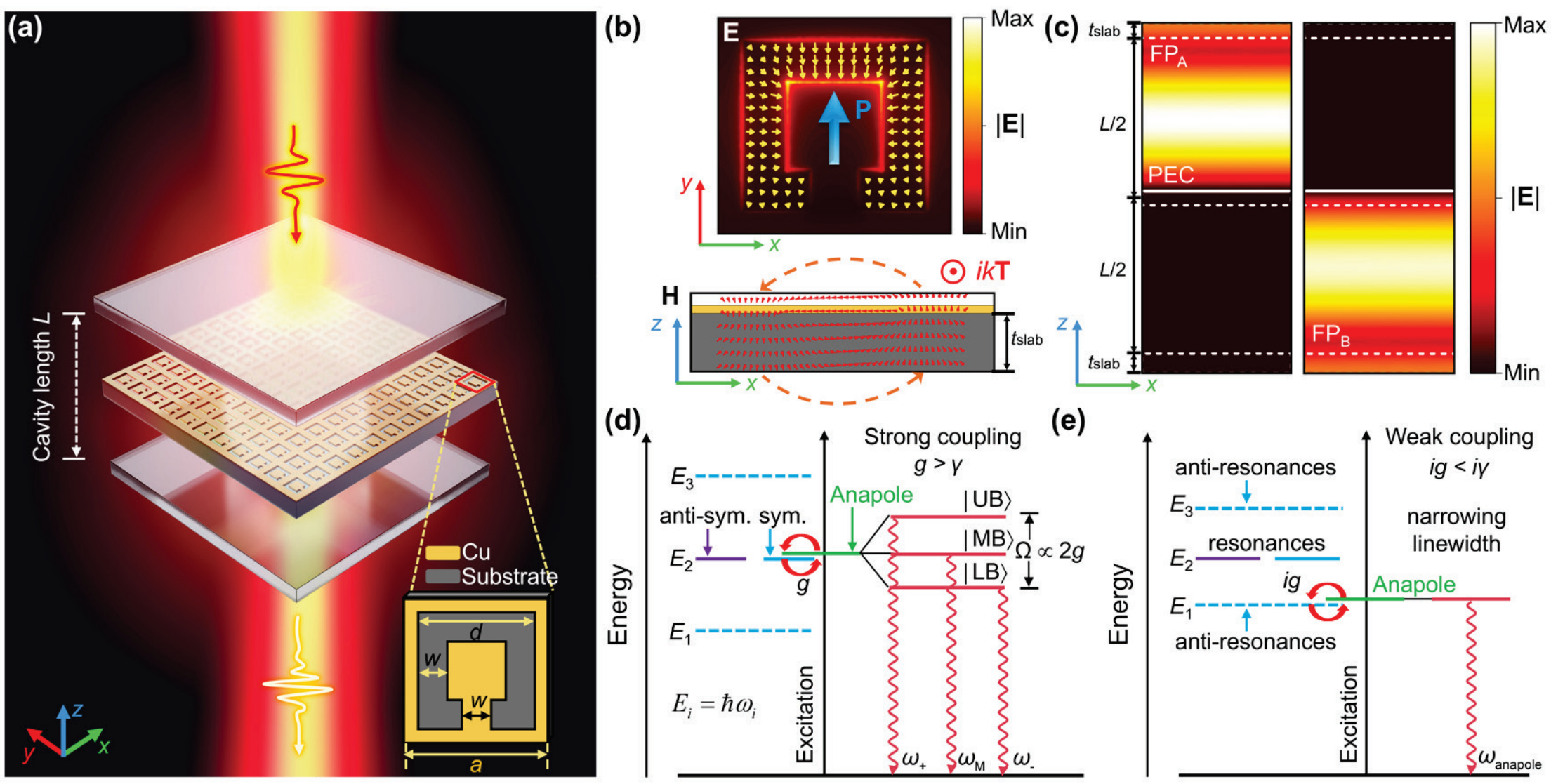

**Figure 1. Structural design and modes evolution of the hybrid cavity.** (a) Schematic of the hybrid cavity constituted by an anapole metasurface inside a Fabry-Pérot (FP) cavity. Inset: unit cell of the metasurface with labeled parameters, including side length $d$ = 0.4 mm, and width of the slot or the gap $w$ = 0.1 mm. (b) Normalized electric field (|**E**|) distribution of the metasurface's lowest eigenmode reveals a $y$-aligned electric dipole moment (**P**). Magnetic field (**H**) distribution indicates a toroidal dipole moment ($ik$**T**) also oriented along the $y$ direction. They are out of phase and form an anapole. The substrate thickness $t_{\mathrm{slab}}$ is 0.13 mm. (c) Metallic middle layer instead of metasurface pattern isolates the full FP cavity into two separate FP modes ($FP_A$ and $FP_B$) concentrated in the dielectric-metal regions. The white solid lines and dashed lines delineate the perfect electric conductor (PEC) and dielectric slabs, respectively. (d, e) Diagrams illustrate the coupling between the anapole and FP cavity's energy levels, including strong coupling with resonances (d), and weak coupling with anti-resonances (e).

To probe the interaction within the hybrid cavity, we simulate and map its transmission under $y$-polarized incidence across the cavity length $L$ and frequency $f$ [Fig. 2(a)]. The response exhibits distinct phenomena spanning from linewidth narrowing to Rabi splitting. Meanwhile, semi-analytical calculations with the transfer matrix method (TMM) validate all the characteristics (see detailed calculation process in Note S5) and show excellent agreement with numerical simulations in the considered range. At $L$ = 2.75 mm, three transmission peaks emerge [Fig. 2(b)], signifying coherent energy exchange and strong mode hybridization between the anapole and two separated FP cavities. While at $L$ = 1.3 mm, the transmission of the hybrid cavity only exhibits a single and narrowband resonance peak [Fig. 2(d)], with its linewidth much narrower compared to that of the bare metasurface. In fact, the free-standing placement of the metasurface plays an essential role in inducing these responses. If the metasurface were used as a mirror to constitute the FP cavity instead of freestanding in the middle, the coupling dynamics would differ substantially: MB would vanish, while linewidth narrowing and peak intensity would be strongly suppressed (see Note S6 in Supporting Information). Going back to our case, for further analysis, we analytically extend the transmission of the hybrid cavity from TMM to the complex-frequency plane. When $L$ = 2.75 mm, three poles are observed around the anapole frequency, corresponding to the two polaritonic branches and an anti-symmetric cavity mode [Fig. 2(c)], while at $L$ = 1.3 mm, there is only a single pole at the anapole frequency with significantly decreased Im($f$) [Fig. 2(e)]. The real and imaginary parts of these poles calculated from TMM agree excellently with the counterparts of the eigenmodes retrieved from simulations.

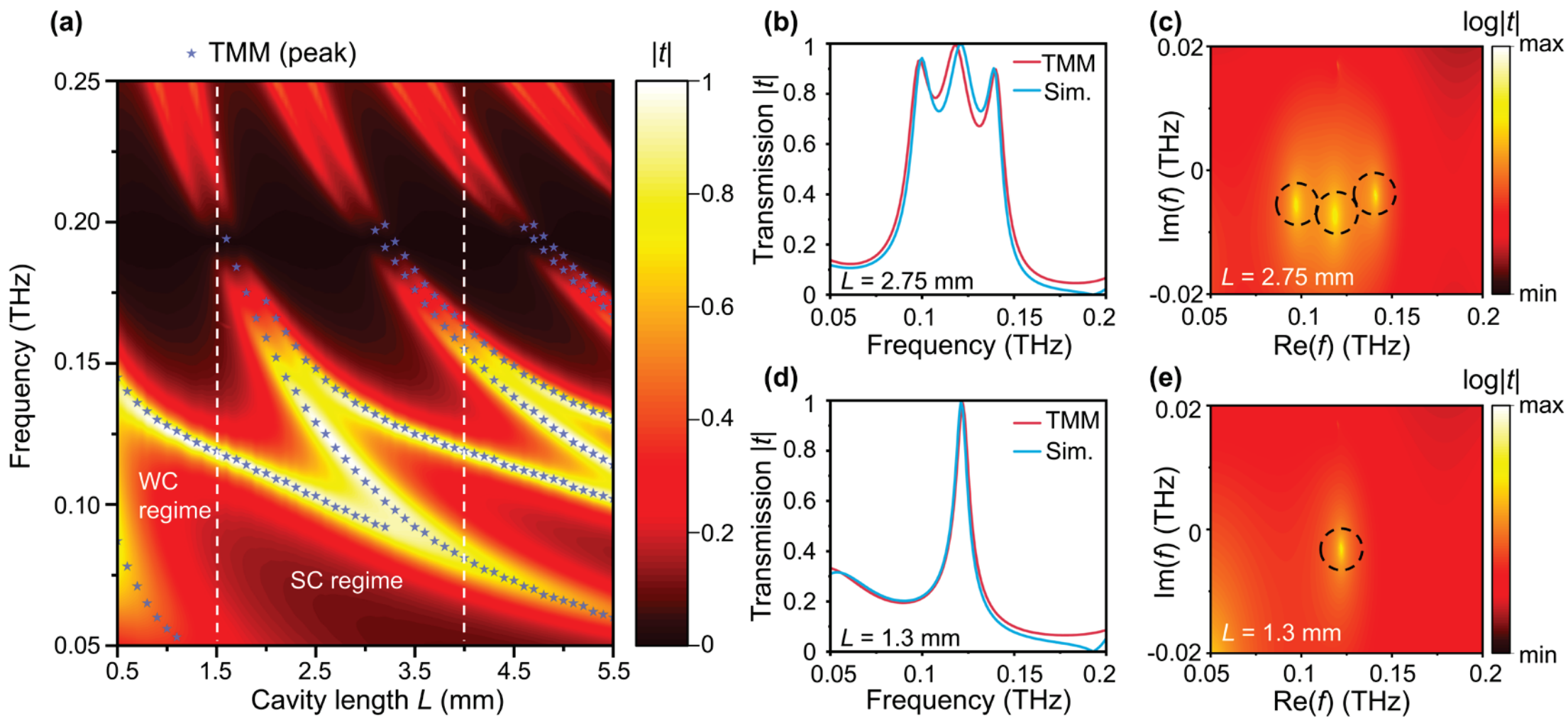

**Figure 2. Numerical simulations and theoretical analysis of the hybrid cavity.** (a) The transmission color map of the hybrid cavity for varying the cavity length $L$, ranging from 0.5 to 5.5 mm. It is delineated into weak and strong coupling regimes according to coupling strength. Purple stars indicate transmission peaks obtained using the transfer matrix method (TMM). (b, d) The transmission spectra calculated from the simulations and TMM for $L$ = 2.75 mm (b) and 1.3 mm (d), respectively. (c, e) The complex-frequency transmission maps obtained via TMM for $L$ = 2.75 mm (c) and 1.3 mm (e), respectively. Dashed circles denote the poles that indicate the eigenmodes after coupling.

## 2.2 Strong Coupling with Resonances

To elucidate the underlying physics, we systematically investigate coupling properties. Anapole mode of the metasurface exhibits a transmission peak at 0.12 THz with a full-width half-maximum (FWHM) of 34 GHz [Fig. 3(a)]. Tuning the cavity length to $L$ = 2.75 mm, which spectrally overlaps the 2nd-order FP mode ($E_2$) with the anapole, three transmission peaks (LB, MB, and UB) are generated after coupling, experimentally verified via THz time-domain spectroscopy (TDS) [Fig. 3(b); see schematic illustration of setup in Note S7]. LB and UB are attributed to strong-coupling-induced Rabi splitting in analogy to the polaritonic branches, while MB exhibits dispersion behavior and antiparallel far-field radiation similar to the anti-symmetric FP mode (see Note S8 for details). For further investigation, we adopt the COM to describe the coherent interaction in the strong coupling regime quantitatively[49,50]:

$$\begin{pmatrix} \omega_{\text{anapole}} - i\gamma_{\text{anapole}} & 0 & g \\ 0 & \omega_{\text{anti-sym.FP}} - i\gamma_{\text{anti-sym.FP}} & 0 \\ g & 0 & \omega_{\text{sym.FP}} - i\gamma_{\text{sym.FP}} \end{pmatrix} \begin{pmatrix} V_1 \\ V_2 \\ V_3 \end{pmatrix} = \omega_{\text{H}} \begin{pmatrix} V_1 \\ V_2 \\ V_3 \end{pmatrix}, \quad (1)$$

in which the angular frequencies and dampings of anapole ($\omega_{\text{anapole}}$ and $\gamma_{\text{anapole}}$) and FP modes ($\omega_{\text{FP}}$ and $\gamma_{\text{FP}}$) are directly extracted from simulations. $\omega_{\text{H}}$ corresponds to the frequencies of hybridized eigenstates. Eigenvectors $(V_1, V_2, V_3)^{\text{T}}$ are determined to quantify contributions from the anapole ($|V_1|^2$), anti-symmetric FP ($|V_2|^2$), and symmetric FP modes ($|V_3|^2$), respectively, with the condition: $|V_1|^2 + |V_2|^2 + |V_3|^2 = 1$. Due to the strong coupling that occurs exclusively between the anapole and symmetric FP mode, $g$ appears only in the anti-diagonal terms of the Hamiltonian matrix. Through diagonalization of the Hamiltonian matrix Eq. (1), the dispersion relations are calculated as follows[22]:

$$\begin{aligned} \omega_{\pm} &= \frac{(\omega_{\text{anapole}} - \gamma_{\text{anapole}} + \omega_{\text{sym.FP}} - \gamma_{\text{sym.FP}})}{2} \pm \sqrt{g^2 - \frac{1}{4}[i\delta + (\gamma_{\text{anapole}} - \gamma_{\text{sym.FP}})]^2}, \\ \omega_{\text{M}} &= \omega_{\text{anti-sym.FP}}. \end{aligned} \quad (2)$$

With the above parameters, the dispersion relations are plotted as cavity length versus frequency in Fig. 3(c). The analytically derived results using COM show excellent consistency with numerical simulations. At $\delta = 0$, two branches asymptotic to the anapole frequency are disconnected by a polaritonic gap, characterized by anti-crossing behavior with a Rabi splitting of ~41.3 GHz (> $\gamma$ = 24 GHz). And the normalized coupling ratio $\Omega_{\text{Rabi}}/\omega_{\text{anapole}}$ = 34.4% exceeds the strong coupling category. These strict criteria confirm that our hybrid cavity is indeed within the ultrastrong coupling regime[48]. In addition, coherent interactions between an anapole and higher-order cavity resonances reveal smaller frequency separations, scaling as $g \propto \sqrt{1/V}$, where $V$ is the mode volume (see Note S9 in Supporting Information). By maintaining a constant dielectric environment within the cavity, the coupling strength can be flexibly tuned and further increased by varying the slab thickness and permittivity (see Note S10 for further discussion).

Akin to strong coupling of light-matter interaction[51], the anapole and cavity modes equally contribute ($|V_1|^2 = |V_3|^2 = 0.5$) to the LB and UB at the zero detuning energy, as shown in Fig. 3(d). Within the range ($L$ < 2.75 mm), LB are dominated by the anapole mode in terms of $|V_1|^2 > 0.5$ (anapole-like), and they are otherwise cavity-like outside this range. Figure 3(e) shows LB's mode volume $V$ increasing with cavity length $L$,

revealing its mode characteristics from anapole-like to cavity-like transition, while UB decreases inversely. MB consistently dominated by the anti-symmetric FP mode ($|V_2|^2 = 1$), resulting in a mode volume trend in the strong coupling regime similar to that of the 2nd-order cavity mode. Additionally, with the tuning of the cavity length $L$ that determines the mode fraction of the LB and UB, not only the mode volume, but the damping of the LB and UB also cross each other[50], while the MB's damping remains nearly unchanged (see Note S11 in Supporting Information). In fact, the polaritons outperform the original anapole and FP modes for light-matter interaction, combining their advantages while providing additional benefits. For example, they provide one order of magnitude smaller mode volumes compared to FP cavities, along with a tunable operating frequency range that corresponds to the Rabi splitting frequency of ~41.3 GHz, which exceeds 30% of the anapole frequency ($\Omega_{\text{Rabi}}/\omega_{\text{anapole}}$ = 34.4%). Moreover, we calculate the projected LDOS under $y$-polarized excitation near the metasurface (see Eq. S12 in Note S12). The LDOS spectra reveal a significant enhancement in the hybrid cavity that substantially exceeds the bare metasurface response [see Fig. 3(f) for quantification of LB and UB], especially for the UB, which has the smallest mode volume when $L$ = 2.75 mm. In addition, the LDOS spectrum of the MB is basically consistent with that of a FP cavity mode (see Note S13 in Supporting Information), further demonstrating that it is isolated from hybridization.

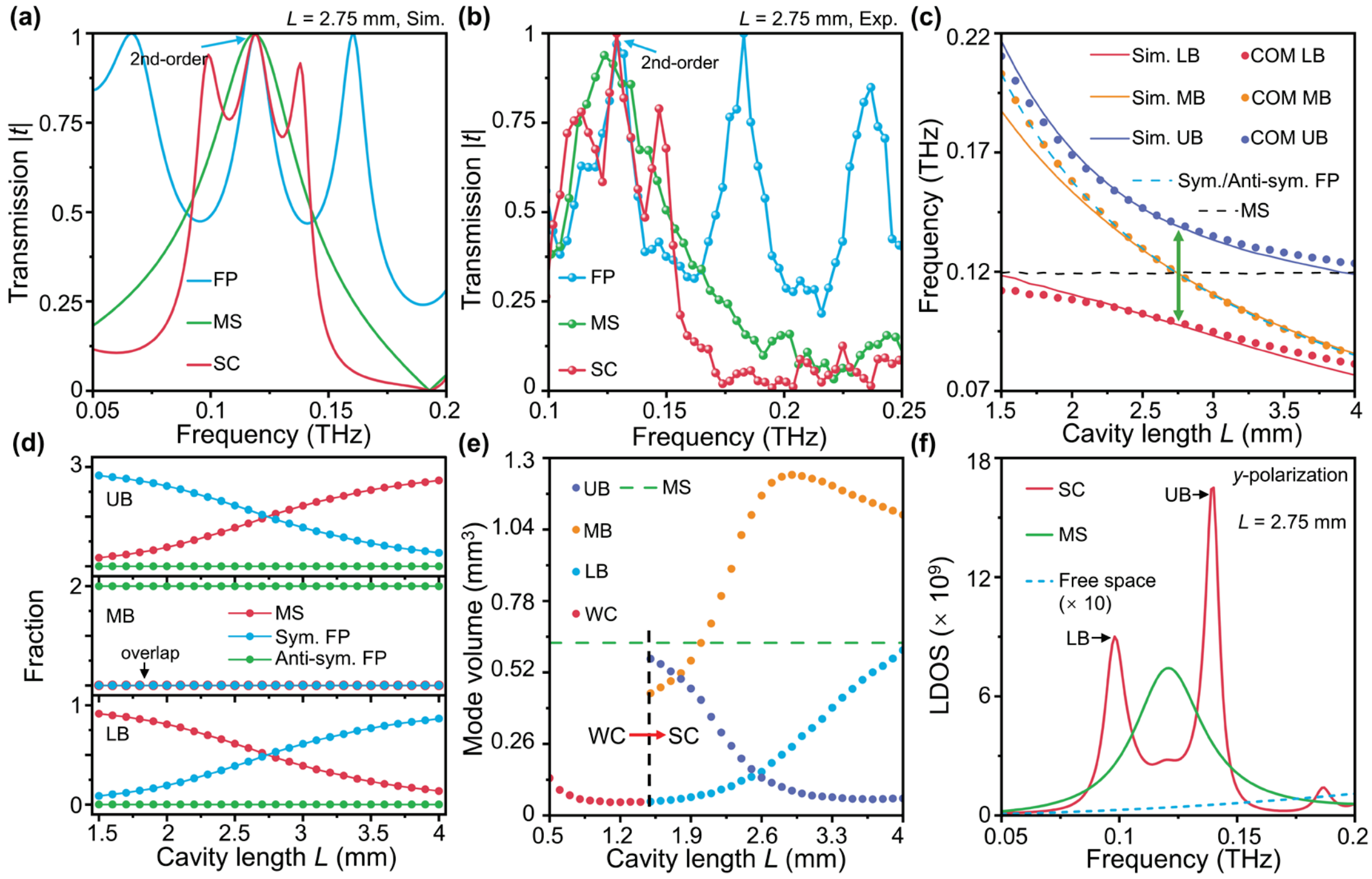

**Figure 3. Strong coupling between anapole and FP cavity resonances.** (a, b) Simulated (a) and experimental (b) transmission spectra for the hybrid cavity with $L$ = 2.75 mm in the strong coupling (SC) regime. Blue and green lines: transmission spectra of the 2nd-order FP and bare metasurface (MS) modes. (c) Dispersions of the three coupled modes versus the cavity length $L$. The numerical results are denoted by red (LB), orange (MB), and purple lines (UB), exhibiting excellent agreement with the results obtained from the COM (dots), with a Rabi splitting frequency ~41.3 GHz (green arrow). Black dashed line: bare metasurface (MS).

Blue dashed line: near-degenerate symmetric/anti-symmetric FP modes. (d) Fractions of the bare metasurface, symmetric FP, and anti-symmetric FP modes in the LB, MB, and UB, respectively. (e) Mode volume for eigenmodes transitioning from weak to strong coupling regimes. (f) Local density of states (LDOS) near the metasurface inside the FP cavity in their strong coupling ($L$ = 2.75 mm) compared to that of the bare metasurface.

### 2.3 Weak Coupling with Anti-resonances

As shown in Fig. 4(a), the hybrid cavity at $L$ = 1.3 mm exhibits a much narrower transmission peak with a FWHM of 6 GHz (red line), representing a nearly 5.7-fold reduction relative to the bare metasurface's (green line). Experimentally, we also observe the remarkable spectral narrowing of the transmission peak [Fig. 4(b)], despite a reduced peak value as the frequency resolution is inherently constrained in the THz TDS (optimally ~2.5 GHz for our delay line)[52]. We note that this much narrower transmission peak due to weak coupling still forms an anapole at resonance (see Note S14 in Supporting Information for verification). To understand this phenomenon, we quantify the intracavity energy distribution using the energy fraction $\eta$, defined as $\int_{V_{\mathrm{cav}}} W(r)d^3r / \int_V W(r)d^3r$, where $V_{\mathrm{cav}}$ and $V$ denote the volume of the hybrid cavity and the entire computational domain, respectively, and $W(r)$ is the electromagnetic energy density[53] (see Note S15 for details). We consider the three cases: the bare metasurface, the separated FP cavity, and the hybrid cavity [Fig. 4(c)]. For the separated cavity, excited at a frequency corresponding to the 1st-order FP mode ($E_1$), its $|\mathbf{E}|$ distribution reveals pronounced energy exclusion from the cavity interior [Fig. 4(d)], which is a typical anti-resonant phenomenon[54]. This observation highlights that the weak coupling arises from the interaction between the anapole and the anti-resonances, which strongly "squeeze" the anapole mode into the split-ring slots, amplifying the field around it [Fig. 4(e)]. This localization, after coupling, thereby hinders the radiation and extends the photon lifetime within the cavity. In essence, this linewidth narrowing is a direct result of weak coupling between the anapole and anti-resonances, which can be represented by a pure imaginary number[55]. Notably, the coupling strength $ig$ is related to the damping of the anapole mode, as revealed by TMM calculations (see Note S16 in Supporting Information for details). Although higher-order FP anti-resonances can further compress the linewidth, their smaller free spectral range causes the transmission to be no longer a single band [see Fig. S14(a) in Note S12]. Along with the amplified field strength, a substantial enhancement of $y$-polarized projected LDOS near the metasurface is achieved [Fig. 4(f)]. For the weak-coupling regime ($D_{\mathrm{WC}}$) and its counterpart without cavity ($D_{\mathrm{MS}}$), both LDOS spectra follow the Lorentzian lineshapes[56,57]:

$$D(f) = \frac{q}{2\pi^2} \frac{f_0 / 2Q}{(f - f_0)^2 + (f_0 / 2Q)^2}. \tag{3}$$

Here, $f_0$ is the anapole resonant frequency, and $q$ is a fitting parameter. Interestingly, quality factors derived from the projected LDOS for the bare metasurface ($Q_{\mathrm{MS}}$) and the hybrid cavity ($Q_{wc}$) are comparable to those observed in transmission spectra. The change in the projected LDOS is characterized by the Purcell factor $F_p$, defined by the ratio $D_{wc}/D_{MS}$. At the resonance with $f = f_0$, $F_p$ can be directly simplified to $Q_{wc}/Q_{MS}$. The $Q_{MS}$, obtained from the fitting results, is approximately 3.7. In contrast, $Q_{wc}$ is 18.8, resulting in a Purcell factor of ~5.1, which agrees with the linewidth compression ratio. We note that the enhancement of the $Q$ factor is due to the squeezing of the anapole mode by anti-resonances, which also significantly enhances the photonic LDOS. Thus, this enhancement is in stark contrast to previous studies of a single quantum well in a bandgap of photonic crystals, where the increase of emission lifetime is essentially due to the inhibition of in-plane radiation induced by near-zero photonic LDOS[58,59]. We also emphasize that, in contrast to conventional resonance-matching methods[60], the LDOS enhancement observed in our hybrid cavity results directly from spatial LDOS localization (see Note S17 in Supporting Information). This localization in our hybrid cavity, or mode volume compression [Fig. 3(e)], is caused by anti-resonances of the FP cavities. Therefore, to achieve higher LDOS enhancement, we increase the permittivity of the dielectric slabs that form the FP cavity, thereby boosting its $Q$ factor to enhance the Purcell effect. We also decrease the slab thickness accordingly in order to simultaneously fulfill the thin-slab condition ($< \lambda_{slab}/10$) that avoids mode leakage due to interference in the dielectric slabs. This simple optimization yields a two-order-of-magnitude LDOS enhancement in the hybrid cavity compared to that of the bare metasurface (see Note S18). Such an enormous LDOS enhancement enables our hybrid cavity to have outstanding performance in many potential applications. For example, with enhanced spectral selectivity and a signal-to-noise ratio, we ultimately demonstrate a 5.7-fold improvement in the figure of merit in refractive index sensing (see Notes S19 and S20 in Supporting Information).

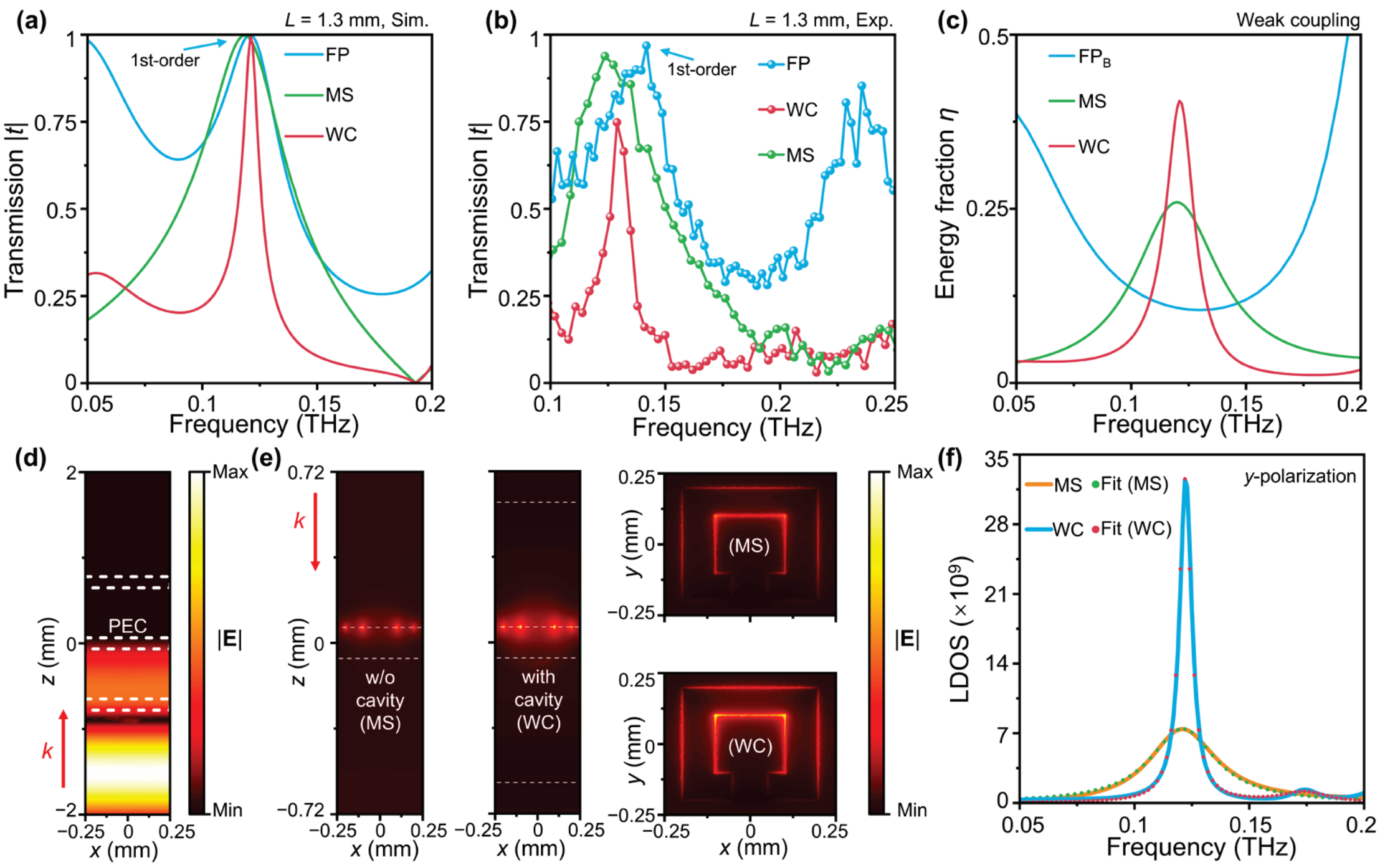

**Figure 4. Weak coupling between anapole and FP anti-resonances.** (a, b) Simulated (a)

and experimental (b) transmission spectra for the hybrid cavity ($L$ = 1.3 mm), compared to bare metasurface (green lines) and 1st-order FP (blue lines) spectra. (c) Calculated energy fraction $\eta$ versus frequency for $FP_B$, bare metasurface, and hybrid cavity (WC), respectively. (d) |**E**| distribution exhibiting the FP anti-resonance, which strongly inhibits the excitation inside the cavity. (e) |**E**| distributions: bare metasurface (MS); hybrid cavity in the weak coupling (WC). Both *x*-*z* plane and *x*-*y* plane distributions are shown. Red arrows indicate the direction of the incident. (f) LDOS comparison between the hybrid cavity and bare metasurface, and their fitting results via the Lorentz model.

Finally, we examine the coupling in the hybrid cavity when the metasurface is positioned away from the center of the full cavity. As noted earlier, when centered, the metasurface coincides with the anti-nodes (nodes) of odd-order (even-order) FP modes, and the metallic layer will convert these FP modes into anti-resonances (resonances) of the half cavities. However, displacing the metasurface from the center alters this correspondence. Therefore, the coupling within our hybrid cavity can be switched via only passive control. While active materials, such as liquid crystals and pyroelectric materials, have been investigated for various similar applications[61–63], our design establishes an intrinsic mechanism for modification and switch of coupling without any dependence on external excitation conditions. With our hybrid cavity, this mechanism enables direct manipulation of its dispersion, coupling strength, and LDOS. To elucidate this point, two specific configurations involving high-order FP modes are considered [Fig. 5(a)]: metasurface at (i) an anti-node of the 2nd-order FP mode ($E_2$), and (ii) a node of the 3rd-order FP mode ($E_3$). We decompose the coupling of the metasurface to the FP cavity into its interaction with the two individual half-cavities ($FP_A$ and $FP_B$) for a comprehensive investigation. In configuration (i), spectral overlap between the anapole resonance and anti-resonance ranges of the $FP_A$ and $FP_B$ induces weak coupling (see Note S21 in Supporting Information for further verification), resulting in narrowing transmission linewidths of 18.7 and 14 GHz, respectively [Fig. 5(b, c)]. The observed reduction in peak transmission intensity is attributed to impedance mismatch at the interface, which diminishes coupling efficiency and overall system performance, even in cases of strong coupling. In contrast to a hybrid cavity, integrating the metasurface with both half-cavities simultaneously exhibits a narrower transmission linewidth (5.3 GHz) and higher transmission intensity [see Fig. 5(f) for comparison]. This enhanced performance achieves more effective suppression of radiation loss and a concomitant extension of the photon lifetime within the full cavity. Conversely, configuration (ii) features spectral overlaps between the anapole resonance frequency and the resonant frequencies of both $FP_A$ and $FP_B$, which drives the hybrid cavity into the strong coupling regime (see Note S22 for details). Notably, when the metasurface interacts with the half-cavities individually, only two polaritonic branches emerge [Fig. 5(d, e)]. The coupling strength between the anapole mode and $FP_B$ is weaker than that observed with $FP_A$, consistent with $FP_B$'s larger mode volume, as discussed previously. Although combining both half-cavities (3rd-order resonance) leads to an even larger mode volume than the case where they are independent (1st-order and 2nd-order resonances), the entire hybrid cavity satisfies the impedance matching condition, enabling enhanced coupling strength and improved multi-dispersion tunability in tripartite coupling [see Fig. 5(f)]. These advantageous characteristics are inherently difficult to achieve in a simplified system comprising only a single half-cavity coupled to the anapole metasurface, which also means that the metasurface should be placed inside the cavity rather than at the boundary. In general, when the metasurface is placed at an arbitrary position within the cavity, the anapole metasurface will selectively couple with only a single half cavity at resonance instead of anti-resonance, and the properties will deteriorate accordingly (see Note S23 in Supporting Information for example). Therefore, the metallic metasurface's placement at nodes or anti-nodes within the cavity determines the correspondence rules for specific FP modes to act as either resonances or anti-resonances that couple with the anapole, thereby achieving optimal performance.

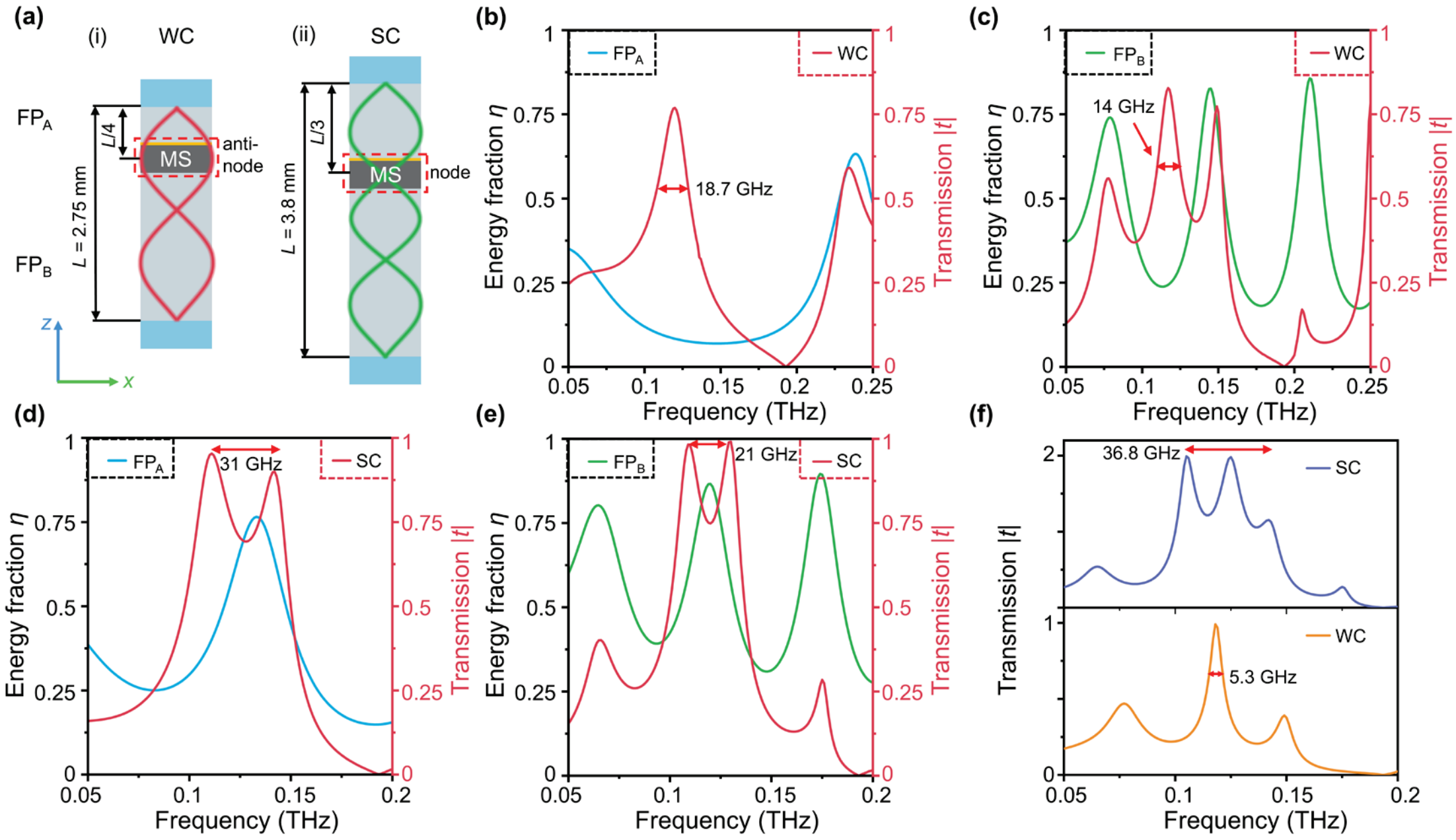

**Figure 5. Position-switchable resonances or anti-resonances coupling with an anapole.** (a) Schematic representations of the metasurface (MS) placed at different spatial positions inside the FP cavity: positioned metasurface at anti-node enables weak coupling (WC) to FP anti-resonances of half cavities (b, c), and at node enables strong coupling (SC) to FP resonances of half cavities (d, e), respectively. Blue curves in (b, d) and green curves in (c, e) represent energy fractions of the $FP_A$ and $FP_B$ cavities, respectively, indicating their anti-resonances (dip) or resonances (peak). (f) The transmission spectra of the hybrid cavity under WC (bottom spectrum) and SC regimes (top spectrum), corresponding to the configurations in (a), exhibit a narrower linewidth and larger Rabi splitting, respectively.

## 3 Conclusion

In summary, we propose and experimentally demonstrate a tunable hybrid cavity integrating an anapole metasurface inside an FP cavity, enabling switching between strong and weak coupling regimes via cavity-length control. The hybrid cavity can enable precise repositioning of the metasurface to specific locations, such as the node or anti-node of the full cavity mode, resulting in the half cavities simultaneously exhibiting resonances or anti-resonances, respectively. This way introduces an extra dimension of tunability for achieving the coupling transition. On the one hand, in the strong coupling regime between the anapole and FP resonances, the hybrid cavity achieves a normalized coupling ratio $\Omega_{\mathrm{Rabi}}/\omega_{\mathrm{anapole}}$ ~34.4%, thus entering the ultrastrong coupling regime. The generated polaritons combine the favorable properties of both anapole and cavity resonances, allowing tunable resonant frequencies while maintaining substantially tighter mode confinement. Our hybrid cavity overcomes the inherent constraints of weak coupling strength due to large mode volumes in long wavelengths, opening unprecedented opportunities for enhancing light-matter interaction efficiency. On the other hand, in the weak coupling regime, we achieve a 5.7-fold narrowing of the single-band linewidth by spectral aligning of the anapole with FP anti-resonances, establishing a novel strategy that departs from conventional cavity resonance-induced damping reduction. Optimizing high-$Q$ FP cavities boosts the Purcell effect, with the hybrid cavity delivering two orders of magnitude LDOS enhancement via this mechanism. This underlying scheme for realizing tunable coupling is readily extendable to optical frequencies (e.g., anapole silver metasurface integrates with a silicon-based cavity, see Note S24). These findings demonstrate that the tunable hybrid cavity provides a versatile platform for photons to couple with active or nonlinear materials, especially in long wavelengths, thereby promoting the development of dynamically reconfigurable hybrid devices for advanced photonics and quantum information processing. Further integrated with emerging quantum materials[64,65], it could also enable coherent and cavity-tunable manipulation of charge-transfer dynamics, phonon-polariton transport, and vibro-polariton chemistry through tailored strong-interaction environments.

## Supplemental Material

See SI for supporting content.

## Acknowledgment

This research was supported by the National Natural Science Foundation of China (No. 12304348), Guangdong Basic and Applied Basic Research Foundation (No. 2025A1515011470), Guangdong University Featured Innovation Program Project (2024KTSCX036), Guangdong Provincial Project (2023QN10X059), Guangzhou-HKUST(GZ) Joint Funding Program (2025A03J3783), Guangzhou Municipal Science and Technology Project (2024A04J4351), Guangzhou Young Doctoral Startup Funding (2024312028).

## Author Competing Interest

The authors declare no conflict of interest.